\documentclass{KapProc} % Computer Modern font calls
\usepackage{psfig}

\let\footnote\savefootnote
\let\footnotetext\savefootnotetext 
\let\footnoterule\savefootnoterule 
\normallatexbib
\def\be{\begin{equation}}
\def\ee{\end{equation}}
\def\bea{\begin{eqnarray}}
\def\eea{\end{eqnarray}}

\begin{document}

\articletitle[THE PAST AND FUTURE OF COULOMB DISSOCIATION
%IN HADRON- AND ASTROPHYSICS
]
{The Past and Future of Coulomb Dissociation
in Hadron- and \\Astrophysics}

\author{G. Baur{\dag}, K. Hencken{\ddag}, D. Trautmann{\ddag}, S. Typel{\S} and H. H. Wolter{\S}}

%% affil, email, and abstract are optional
\affil{{\dag} Forschungszentrum J\"ulich\\
       Institut f\"ur Kernphysik\\ 
       D-52425 J\"ulich\\
       Germany}
%\footnote{Partial funding provided by grant NL-213-456.}}
\email{g.baur@fz-juelich.de}

\affil{{\ddag} Institut f\"ur Theoretische Physik\\
       Universit\"at Basel\\
       Klingelbergstra\ss{}e 82\\
       CH-4056 Basel\\
       Switzerland}
\email{k.hencken@unibas.ch}
\email{dirk.trautmann@unibas.ch}

\affil{{\S} Sektion Physik\\
       Universit\"at M\"unchen\\
       D-85748 Garching\\
       Germany}
\email{stefan.typel@physik.uni-muenchen.de}
\email{hermann.wolter@physik.uni-muenchen.de}

%% optional, to supply a shorter version of the title for the running head:
%%\chaptitlerunninghead{}

% please insert the keywords of your article below

\begin{abstract}
Breakup reactions are generally quite complicated, they involve
nuclear and electromagnetic forces including interference effects.
Coulomb dissociation is an especially simple and
important mechanism since  the
 perturbation due to the electric field 
of the nucleus is exactly known. Therefore firm conclusions can be drawn
from such measurements. Electromagnetic matrixelements, radiative capture 
cross-sections and astrophysical S-factors can be extracted from experiments. 
We describe the basic theory, give analytical results
for higher order effects in the dissociation of neutron 
halo nuclei and briefly review the experimental results
obtained up to now.
Some new applications of Coulomb dissociation for nuclear astrophysics 
and nuclear structure physics are discussed.
\end{abstract}

\section{Introduction}

One may regard the work of Oppenheimer and Phillips in 1935 \cite{O35,OP35} 
as a starting 
point of the present subject. They tried to explain the preponderance 
of (d,p)-reactions over (d,n)-reactions by a virtual breakup of the 
deuteron in the Coulomb
field of the nucleus before the actual nuclear interaction takes place.
Because of the Coulomb
repulsion of the proton this would explain the dominance of (d,p)-reactions.
In this context, Oppenheimer~\cite{O35} also treated the real breakup of the 
deuteron in the Coulomb
field of a nucleus. In the meantime, the subject has developed considerably.
In addition to the deuteron, many different kinds of projectiles
(ranging from light to heavy ions, including radioactive nuclei)
have been used at incident energies ranging from below the Coulomb barrier 
to medium and up to relativistic energies.
% Figure 1 aus PRep111
%
\begin{figure}[ht]
\begin{center}
\leavevmode\psfig{file=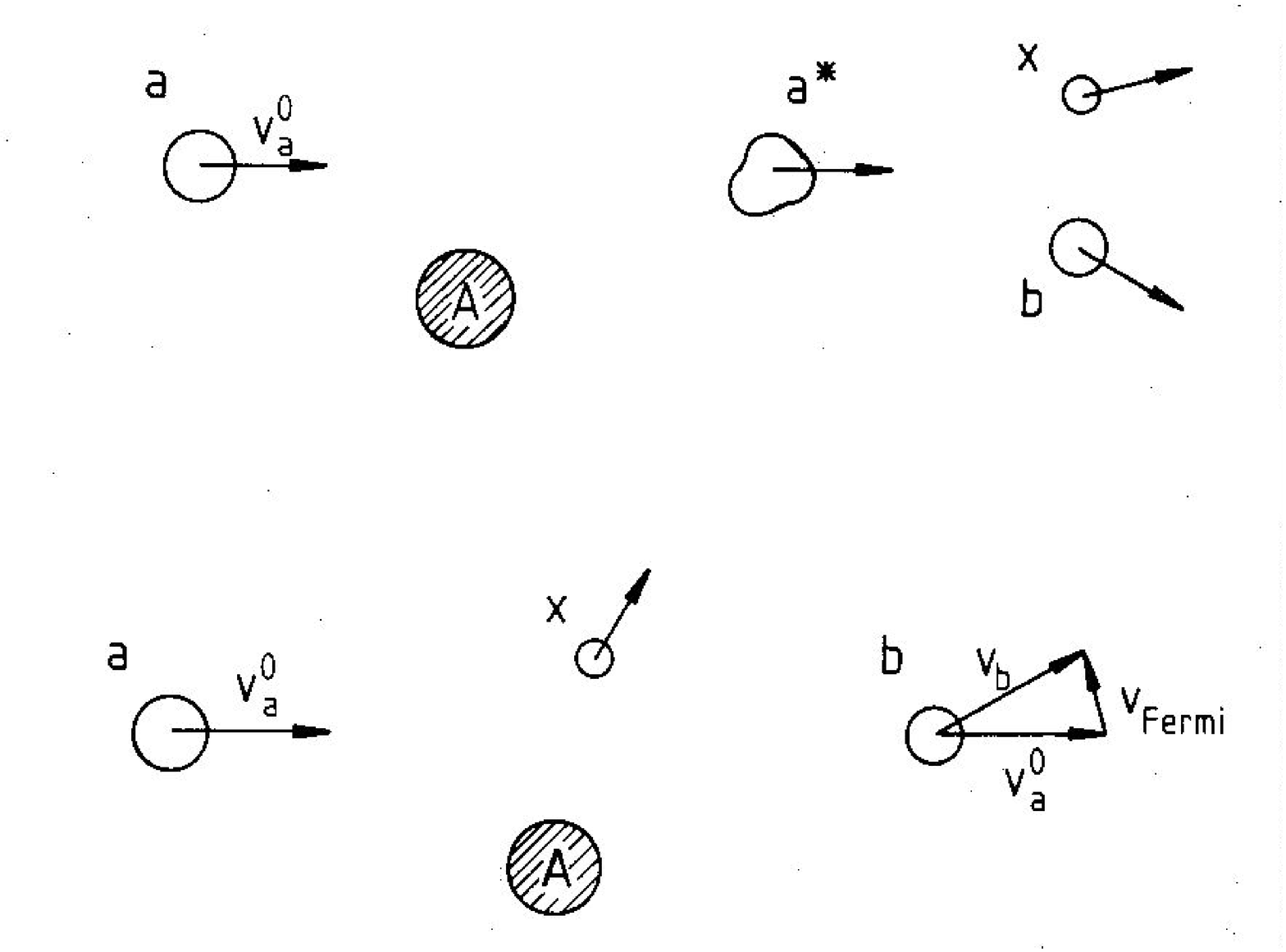,height=6.5cm}
\end{center}
\caption{Two basic reaction mechanisms for breakup are shown
schematically. In the upper figure (sequential breakup), 
the projectile $a$ is excited to a
continuum (resonant) state $a^{\ast}$ which decays subsequently into the
fragments $b$ and $x$. In the lower part (spectator breakup)
substructure $x$ interacts (in
all kinds of ways) with the target nucleus $A$, whereas $b=(a-x)$
misses the target nucleus (``spectator''). It keeps approximately the
velocity which it had before the collision. [Fig.~1 of
Ref.~\protect\cite{BRST84}.]
}
\label{fig_prep1}
\end{figure}

In Fig.~1 we show two different kinds of reaction mechanisms. Since 
rigorous methods of reaction theory (like the Faddeev approach) cannot 
be applied in practice to such complicated nuclear reactions, we have to
use different theoretical methods to treat the different cases as well as 
possible. In  the spectator breakup mechanism the breakup occurs
  due to the interaction of one of the constituents 
with the target nucleus, while the spectator moves on essentially 
undisturbed. 
Another mechanism is the sequential breakup, where the projectile 
is excited to a continuum state which decays subsequently. Both mechanisms have
been dealt with extensively in the past \cite{BRST84}, see also 
Ref.~\cite{TMUPROC},
which provides a brief outline of the development over the last few decades.
For low and medium energies (i.e., for energies not high enough for the Glauber theory to 
be applicable) it should be noted that  
 the post-form DWBA is especially suited to treat 
the spectator  process. For the sequential breakup mechanism
 the decomposition of the Hamiltonian in the inital 
and final channels  is the same.

We start with a general discussion of Coulomb dissociation. Due to the 
time-dependent electromagnetic field  the projectile is excited to a 
bound or continuum state, which can subsequently decay.
We briefly mention the very large effects of electromagnetic excitation
in relativistic heavy ion collisions. 
After a short review of results obtained for nuclear structure
as well as nuclear astrophysics, we discuss new possibilities, like 
the experimental study of two-particle capture.
 We close with conclusions and an outlook.

\section{ Electromagnetic and Nuclear Dissociation}
Coulomb excitation is a very useful tool to determine
nuclear electromagnetic matrixelements. The nuclei are assumed to 
interact with each other only electromagnetically.
 This can be achieved
by either using bombarding energies below the Coulomb barrier or by choosing
very forward scattering angles and high energy collisions. With increasing
beam energy states at higher energies can be excited; this can
lead, in addition to Coulomb excitation, also to Coulomb 
dissociation, for a review see, e.g., Ref.~\cite{Ber01}.
Such investigations are also well suited for secondary (radioactive) beams.
The electromagnetic interaction, which causes the dissociation, is well known
and therefore there can be a clean interpretation of the experimental data. 
This is of
interest for nuclear structure and nuclear 
astrophysics \cite{Bau01,Bau02}. 
Multiple electromagnetic excitation can also be important. We especially
mention two
aspects: It is a way to excite new nuclear states, like the double
phonon giant dipole resonance \cite{Bau02}; 
but it can also be a correction to the
one-photon excitation \cite{Typ01,Typ02,Typ03}.

In the equivalent photon approximation the cross section for an
electromagnetic process is written as
\begin{equation}
 \sigma = \int \frac{d\omega}{\omega} \: n(\omega) \sigma_{\gamma}(\omega)
\end{equation}
where $\sigma_{\gamma}(\omega)$ denotes the appropriate cross section
for the photo-induced process and $n(\omega)$ is  the equivalent
photon number. For sufficiently high  beam energies it is well approximated
by
\begin{equation}
 n(\omega) = \frac{2}{\pi} Z^{2} \alpha \ln \frac{\gamma v}{\omega R}
\end{equation}
where $R$ denotes some cut-off radius. More refined expressions, which
take into account the dependence on multipolarity, beam velocity or
Coulomb-de\-flec\-tion, are available in the 
literature \cite{Ber01,Typ02,Win01}.  
The theory of electromagnetic excitation is well developed for
nonrelativistic, as well as relativistic projectile velocities.
In the latter case an analytical result for all multipolarities was
obtained in Ref.~\cite{Win01}. The projectile motion was treated classically
in a straight-line approximation. On the other hand, in the
Glauber theory, the projectile
motion can be treated quantally \cite{Ber01,Typ03}. 
This gives rise to characteristic diffraction effects. 
The main effect is due to the strong absorption at impact parameters less 
than the sum of the two nuclear radii.

If the above conditions are not met,
nuclear excitation (or diffractive 
dissociation) also has to be taken into account.
This is a broad subject and has been studied in great detail
using Glauber theory, see, e.g., \cite{Ber01} for further references. 
Especially for light nuclei, Coulomb excitation
tends to be less important in general than nuclear 
excitation. For heavy nuclei the situation reverses. 
The nuclear breakup
of halo nuclei was more recently studied, e.g., in  \cite{Hen01}. 
The nuclear interaction of course is less precisely known 
than the Coulomb interaction. In Ref.
\cite{Hen01} the nuclear breakup was studied
 using the eikonal approximation as well as the Glauber
multiple particle scattering theory. 
 No Coulomb interaction was included in this 
approach, as the main focus was on the breakup on light targets. In
Ref. \cite{Mue01} on the other hand, the combined effect of both nuclear 
and Coulomb excitation is studied. The nuclear contribution to the 
excitation is generally found to be small and has an angular dependence 
different from the electromagnetic one. This can be used to separate such 
effects from the electromagnetic excitation.
We also mention the recent systematic study of $^8$B breakup cross
section in \cite{eshe2000}. 

\section{Electromagnetic Excitation in Relativistic
Heavy Ion Collisions} 
Electromagnetic excitation is also used at  relativistic heavy ion
accelerators to obtain nuclear structure information. Recent examples
are the nuclear fission studies of radioactive nuclei \cite{khschmidt00}
and photofission of $^{208}$Pb \cite{abreu99}.
Cross-sections for the excitation of the giant dipole resonance
(``Weizs\"acker-Williams process'')
 at the forthcoming relativistic heavy ion colliders
RHIC  and LHC(Pb-Pb) at CERN are 
huge \cite{RHIC89,Hen02}, of the 
order of 100 b for heavy systems (Au-Au or Pb-Pb).
In colliders,  the effect is considered to be 
mainly a nuisance, the excited particles are lost 
from the beam. On the other hand, the effect will also be useful as a 
luminosity monitor by detecting the neutrons in the forward direction.
Specifically one will measure the  neutrons which will be produced
after the decay of the giant dipole resonance which is excited
in each of the ions (simultaneous excitation).
Since this process has a steeper impact parameter dependence than the 
single excitation cross-section, there is more sensitivity to the 
cut-off radius and to nuclear effects. 
For details and further Refs., see \cite{Hen02}.

\section{Higher Order Effects and Postacceleration}
\label{sec4.2}

Higher order effects can be taken into account in a coupled channels approach,
or by using higher order perturbation theory. The latter involves a sum over
all intermediate states $n$ considered to be important.
Another approach is to integrate the time-dependent 
Schr\"{o}dinger equation directly for a given model Hamiltonian 
\cite{Mel99,Esb01,paris99,typwo99}.
If the collision is sudden, one can neglect the time
ordering in the usual perturbation approach. The interaction can  be summed  
to infinite order. Intermediate states $n$ do not appear explicitly. 

Higher order  effects were recently  
studied in \cite{Tos99}, where further references also to 
 related work can be found. Since full Coulomb
wave functions in the initial and final channels are used there,
the effects of higher order in $\eta_{\rm coul}=\frac{Z Z_c e^2}{\hbar v} $
 are taken into account to all
orders. Expanding the T-matrixelement in this parameter $\eta_{\rm coul}$
one obtains the Born approximation for the dissociation of $a \rightarrow c+n$
\begin{equation}
 T \propto  \frac{\eta_{\rm coul}}{(\vec{q}_n+\vec{q}_c-\vec{q}_a)^2}
 \left(\frac{1}{q_a^2-(\vec{q}_n+\vec{q}_c)^2}
+\frac{1}{q_c^2-(\vec{q}_n-\vec{q}_a)^2}\right).
\end{equation}
This expression is somehow related to  the Bethe-Heitler formula
for brems\-strahlung. The Bethe-Heitler formula has two terms which 
correspond to
a Coulomb interaction of the electron and the target followed by the photon
emission and another one, where the photon is emitted first and then
the electron scatters from the nucleus. In the case of 
Coulomb dissociation  we have a Coulomb scattering
of the incoming particle followed by breakup $a$=$(c+n) \rightarrow c+n$
and another term, where the projectile a breaks up into
c+n, and subsequently, c is scattered on the target. In the case
of bremsstrahlung it is well known \cite{landau} that 
even for $\eta_{coul} \gg 1 $ one obtains the Born approximation result
as long as the scattering is 
into a narrow cone in the forward direction.
This leads one to suspect that higher order effects are not very large
in the case of high energy Coulomb dissociation, when the 
fragments are emitted into the forward direction.

We investigate higher order effects in the model of \cite{Typ01,Typ02,Typ03}.
In a zero range model for the neutron-core interaction, analytical 
results were obtained for $1^{st}$ and $2^{nd}$order electromagnetic excitation
for small values of the adiabaticity parameter $\xi$.
We are especially interested in collisions with small impact 
parameters. For these higher order effects tend to be larger
than for the very distant ones. In this case, the adiabaticity parameter
$\xi$ is small. For $\xi =0$ (sudden approximation)
we have a closed form solution,
where higher order effects are taken into account to all orders. In eq. 37 
of \cite{Typ01} the angle integrated breakup probability is given.
We expand this expression in the strength parameter $\eta_{\rm eff}=
\frac{2 Z Z_c e^2 m_n}{\hbar v (m_n+m_c)}$. We define $x=\frac{q}{\eta}$
where the parameter $\eta$ is related to the binding energy $E_0$
by $E_0=\frac{\hbar^2 \eta^2}{2m}$ and the wave number q is related
to the energy $E_{\rm rel}$ of the continuum final state by $E_{\rm rel}=
\frac{\hbar^2 q^2}{2\mu}$. In leading order (LO) we obtain
\begin{equation}
 \frac{dP_{LO}}{dq}=C \frac{x^4}{(1+x^2)^4}
\end{equation}
where $C=\frac{128 \pi^2 \eta_{\rm eff}^2}{3 \eta^3 b^2}$.
The next to leading order (NLO) expression is proportional to
$\eta_{\rm eff}^4$ and contains a piece from the 2nd order E1 amplitude
and a piece from the interference of 1st and 3rd order. We find
\begin{equation}
 \frac{dP_{NLO}}{dq}=C \left(\frac{\eta_{\rm eff}}{b \eta}\right)^2 
 \frac{x^2 (5-55x^2+28 x^4)}{15 (1+x^2)^6}.
\end{equation}
The integration over $x$
 and the impact parameter $b$ can also be performed analytically in good 
approximation. For details see \cite{tyba}. We can easily 
insert the corresponding values for the Coulomb dissociation 
experiments on $^{11}$Be and $^{19}$C \cite{Nak02,Nak03}
in the present formulae.
We find that the ratio of the NLO contribution to the LO contribution
in the case of Coulomb dissociation on $^{19}$C \cite{Nak03} is given by
$-2\%$. This is to be compared to the results of \cite{Tos99} where a value of
about $-35\%$ was found.

\section{Discussion of some experimental results for nuclear structure and 
astrophysics}
Coulomb dissociation of exotic nuclei is a valuable tool to determine 
electromagnetic
matrix-elements between the ground state and the nuclear continuum. 
The excitation
energy spectrum of the ${}^{10}$Be+n system in the Coulomb dissociation
of the one-neutron
halo nucleus  ${}^{11}$Be on a Pb target at $72\cdot$A~MeV 
was measured \cite{Nak02}.
Low lying E1-strength 
was found. The Coulomb dissociation of the extremely neutron-rich nucleus 
${}^{19}$C was recently studied in a similar way \cite{Nak03}. 
The neutron separation 
energy of ${}^{19}$C could also be determined to be $530\pm130$~keV. 
Quite similarly, the Coulomb
dissociation of the 2n-halo nucleus ${}^{11}$Li was studied in various 
laboratories \cite{Kob89,Shi95,Zin97}. In an experiment at MSU \cite{Iek01}, 
the correlations of the
outgoing neutrons were studied. Within the limits of experimental accuracy,
no correlations were found.

In nuclear astrophysics, radiative capture reactions of the type
$  b + c \to a + \gamma $
play a very important role. They can also be studied in the time-reversed
reaction
 $ \gamma + a \to b + c \: $,
at least in those cases where the nucleus $a$ is in the ground state.
As a photon beam, we use the equivalent photon spectrum which is provided
in the fast peripheral collision. Reviews, both from an experimental
as well as theoretical point of view have been given  \cite{Bau01}, so we want
to concentrate here on a few points.

\begin{figure}[ht]
\begin{center}
\leavevmode\psfig{file=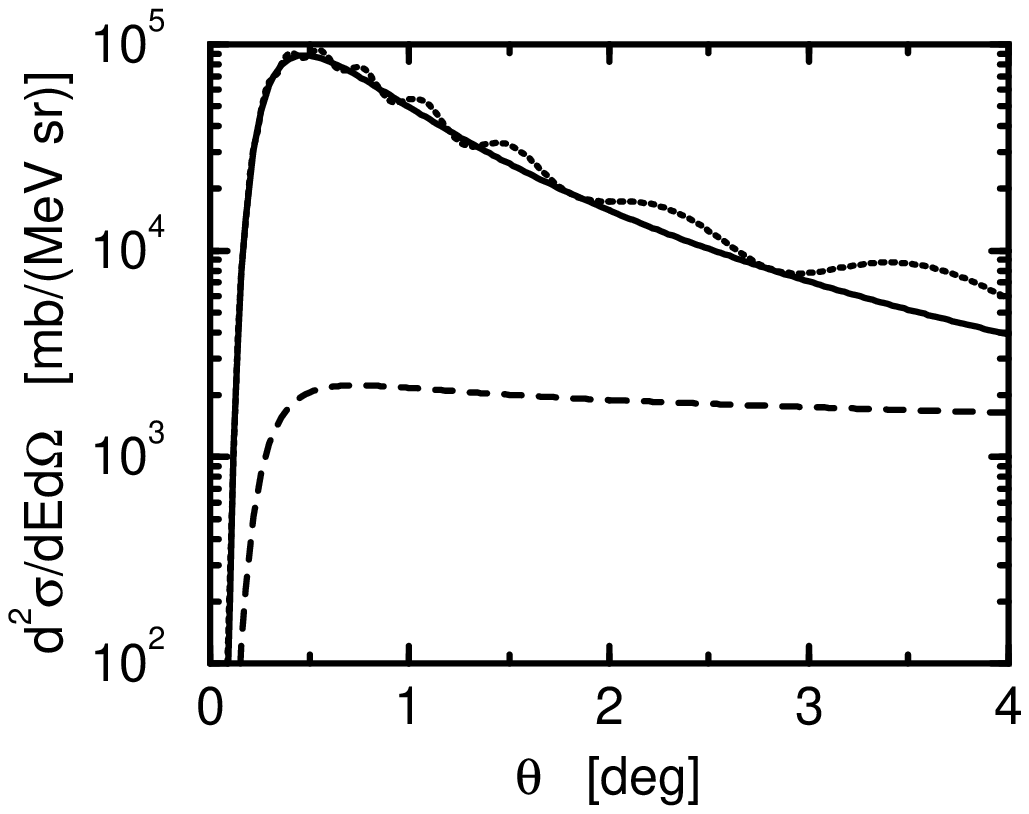,height=5cm,width=5cm}
\leavevmode\psfig{file=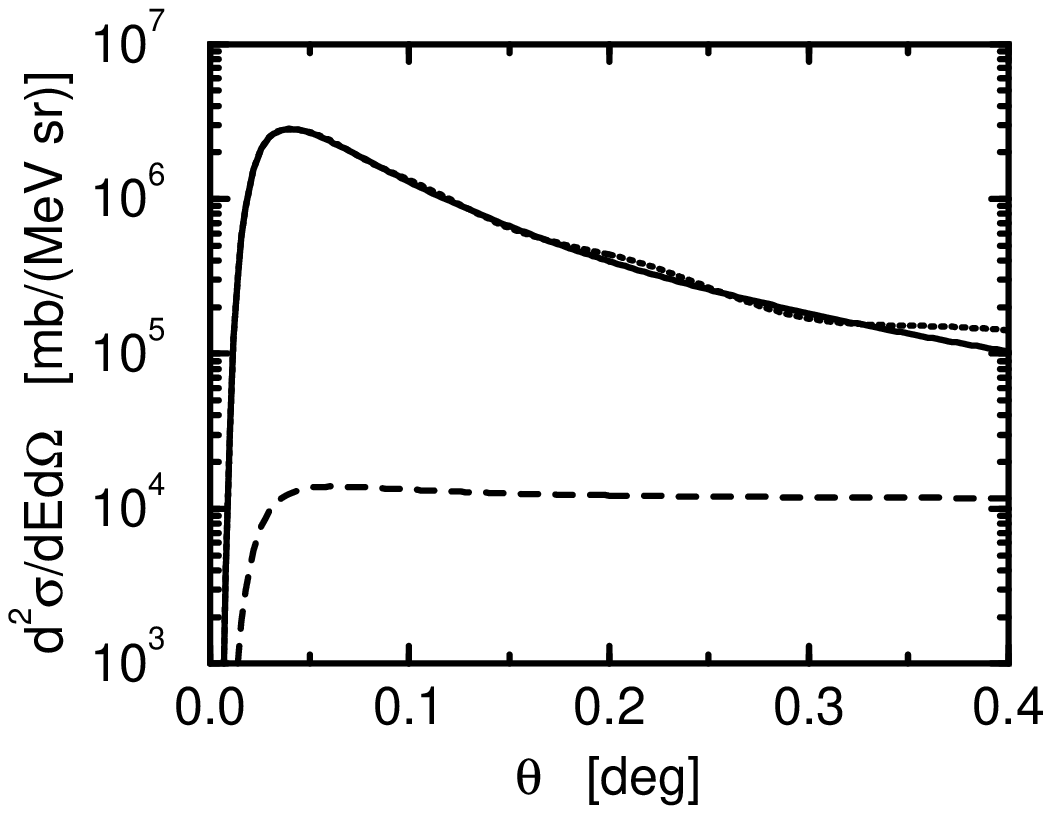,height=5cm,width=5cm}
\end{center}
\caption{Coulomb dissociation cross section of ${}^{8}$B scattered
on ${}^{208}$Pb as a function of the scattering angle for
projectile energies of 46.5~$A\cdot$MeV (left) and 250~$A\cdot$MeV
(right) and a ${}^{7}$Be-p relative energy of 0.3~MeV. 
First order results E1 (solid line), E2 (dashed line) and
E1+E2 excitation including nuclear diffraction (dotted line). 
[From Figs.~4 and 5 of Ref.~\protect\cite{Typ03}.]
} 
\label{procfig}
\end{figure}

The ${}^{6}$Li Coulomb dissociation into $\alpha$+d has been a test case 
of the method, see Ref.~\cite{Bau01}. This is of importance since 
the d$(\alpha,\gamma)^6$Li radiative capture is the only process by
which $^6$Li is produced in standard primordial nucleosynthesis models.
There has been new interest in $^6$Li as a cosmological probe in recent years,
mainly because the sensitivity for searches for $^6$Li has been increasing.
It has been found in metal-poor halo stars at a level exceeding even optimistic 
estimates of how much $^6$Li could have been made in standard big bang 
nucleosynthesis. For more discussion on this see \cite{nollett00}.
  
 The ${}^{7}$Be(p,$\gamma$)${}^{8}$B 
radiative capture reaction is relevant for the solar neutrino problem. 
It determines the
production of ${}^{8}$B which leads to the emission of high energy neutrinos.
There are direct reaction measurements, for a recent one see 
Refs.~\cite{Ham98}.
Coulomb dissociation of ${}^{8}$B
has been studied at RIKEN \cite{Mot02}, MSU \cite{Kel01} and GSI 
\cite{Iwa99}. 
Theoretical calculations are shown in Fig.~2. It is seen that E1 excitation
is large and peaked at very forward angles. E2 excitation is also present,
with a characteristically different angular distribution. Nuclear diffraction
effects are small. Altogether it is quite remarkable that completely 
different experimental methods with possibly
different  systematic errors lead to results that are quite consistent.

\section{Possible New Applications of Coulomb dissociation
 for nuclear astrophysics}

Nucleosynthesis beyond the iron peak proceeds mainly by the r- and s-processes
(rapid and slow neutron capture) \cite{Rol01,Cow01}. 
To establish the quantitative details
of these processes, accurate energy-averaged neutron-capture cross
sections are needed. Such data provide information on the mechanism
of the neutron-capture process and time scales, as well as temperatures
involved in the process. The data should also shed light on neutron
sources, required neutron fluxes and possible sites of the
processes (see Ref.~\cite{Rol01}). The dependence of direct neutron capture
on nuclear structure models was investigated in Ref.~\cite{Rau98}. 
The investigated
models yield capture cross-sections sometimes differing by orders of magnitude.
This may also lead to differences in the predicted astrophysical r-process
paths. Because of low level densities, the compound nucleus model will not 
be applicable.

With the new radioactive beam facilities (either fragment separator or
ISOL-type facilities) some of the nuclei far off the valley of stability,
which are relevant for the r-process, can be produced. In order to assess
the r-process path, it is important to know the nuclear properties like
$\beta$-decay half-lifes and neutron binding energies. Sometimes, the
waiting point approximation \cite{Rol01,Cow01}
is introduced, which assumes an (n,$\gamma$)-
and ($\gamma$,n)-equilibrium in an isotopic chain.
 It is generally believed
that the waiting point approximation should be replaced by dynamic
r-process flow calculations, taking into account (n,$\gamma$), ($\gamma$,n)
and $\beta$-decay rates as well as time-varying temperature and neutron
density. In slow freeze-out scenarios, the knowledge of (n,$\gamma$)
cross sections is important.

In such a situation, the Coulomb dissociation can be a very useful tool
to obtain information on (n,$\gamma$)-reaction cross sections on
unstable nuclei, where direct measurements cannot be done. Of course,
one cannot and need not study the capture cross section on all the nuclei
involved; there will be some key reactions of nuclei close to
magic numbers. It was proposed \cite{Gai01} to use the Coulomb
dissociation method to obtain information about (n,$\gamma$) reaction
cross sections, using nuclei like ${}^{124}$Mo, ${}^{126}$Ru, ${}^{128}$Pd
and ${}^{130}$Cd as projectiles. The optimum choice of beam energy will
depend on the actual neutron binding energy. Since the flux of equivalent
photons has essentially an $\frac{1}{\omega}$ dependence, low neutron
thresholds are favourable for the Coulomb dissociation method. Note
that only information about the (n,$\gamma$) capture reaction to the
ground state is possible with the Coulomb dissociation method. The
situation is reminiscent of the loosely bound neutron-rich light nuclei,
like ${}^{11}$Be, ${}^{11}$Li and ${}^{19}$C.

In Ref.~\cite{Typ01}
the $1^{st}$ and $2^{nd}$ order Coulomb excitation amplitudes
are given analytically in a zero range model for the neutron-core
interaction (see section~\ref{sec4.2}). 
We propose to use the handy formalism of Ref.~\cite{Typ01}
to assess, how far one can go down in beam energy
and still obtain meaningful results with the Coulomb
dissociation method, i.e., where the $1^{st}$ order amplitude
can still be extracted experimentally without being too much disturbed
by corrections due to higher orders. 
For future radioactive
beam facilities, like ISOL od SPIRAL, the maximum beam energy is an
important issue. 
For Coulomb dissociation with two charged particles in the final
state, like in the ${}^{8}$B $\to$ ${}^{7}$Be + p experiment with
a 26~MeV ${}^{8}$B beam  \cite{vSc01} such simple formulae seem to be
unavailable and one should resort to the more involved approaches
mentioned in section \ref{sec4.2}.

A new field of application of the Coulomb dissociation method can be
two nucleon capture reactions. Evidently, they cannot be studied
in a direct way in the laboratory. Sometimes this is not necessary, when the
relevant information about resonances involved can be obtained by
other means (transfer reactions, etc.), like in the triple $\alpha$-process.

Two-neutron capture reactions in supernovae neutrino bubbles are studied
in Ref.~\cite{Goe01}. 
In the case of a high neutron abundance, a sequence of two-neutron
capture reactions, ${}^{4}$He(2n,$\gamma$)${}^{6}$He(2n,$\gamma$)${}^{8}$He
can bridge the $A=5$ and 8 gaps. The ${}^{6}$He and ${}^{8}$He nuclei
may be formed preferentially by two-step resonant processes through their
broad $2^{+}$ first excited states  \cite{Goe01}. Dedicated Coulomb 
dissociation experiments can be useful, see \cite{au99}. Another key reaction can be the
${}^{4}$He($\alpha$n,$\gamma$) reaction \cite{Goe01}. 
The ${}^{9}$Be($\gamma$,n) reaction
has been studied directly (see Ref.~\cite{Ajz01}) 
and the low energy $s_{\frac{1}{2}}$
resonance is clearly established. 

In the rp-process, two-proton capture reactions can bridge the waiting 
points \cite{Bar01,Goe02,Sch02}. From the ${}^{15}$O(2p,$\gamma$)${}^{17}$Ne, 
${}^{18}$Ne(2p,$\gamma$)${}^{20}$Mg and ${}^{38}$Ca(2p,$\gamma$)${}^{40}$Ti
reactions considered in Ref.~\cite{Goe02}, 
the latter can act as an efficient reaction
link at conditions typical for X-ray bursts on neutron stars.
A ${}^{40}$Ti $\to$ p + p + ${}^{38}$Ca Coulomb dissociation experiment
should be feasible. The decay with two protons is expected to be
sequential rather than correlated (``${}^{2}$He''-emission).
The relevant resonances are listed in Table~XII
of Ref.~\cite{Goe02}.
In Ref.~\cite{Sch02} it is found that in X-ray bursts 2p-capture reactions
accelerate the reaction flow into the $Z \geq 36$ region considerably.
In Table~1 of Ref.~\cite{Sch02} nuclei, 
on which 2p-capture reactions may occur,
are listed; the final nuclei are ${}^{68}$Se, ${}^{72}$Kr, ${}^{76}$Sr,
${}^{80}$Zr, ${}^{84}$Mo, ${}^{88}$Ru, ${}^{92}$Pd and ${}^{96}$Cd
(see also Fig.~8 of Ref.~\cite{Bar01}). It is proposed to study the Coulomb
dissociation of these nuclei in order to obtain more direct insight
into the 2p-capture process.

\section{Conclusions}
Peripheral collisions of medium and high energy nuclei (stable or
radioactive) passing each other at distances beyond nuclear contact
and thus dominated by electromagnetic interactions are important tools
of nuclear physics research. The intense source of quasi-real
(or equivalent) photons has opened a wide horizon of related problems
and new experimental possibilities especially for the present and forthcoming 
radioactive beam facilities
to investigate efficiently
photo-interactions with nuclei (single- and multiphoton excitations
and electromagnetic dissociation).

\begin{acknowledgments}
We have enjoyed collaboration and discussions on the present topics with
very many people. We are especially grateful  to 
C.~A.~Bertulani, H.~Rebel, F.~R\"{o}sel, and R.~Shyam.
\end{acknowledgments}

\end{document}